\documentclass[aps,prl,twocolumn,superscriptaddress,showpacs,nofootinbib,longbibliography]{revtex4-1}
\usepackage{graphicx}% Include figure files
\usepackage{dcolumn}% Align table columns on decimal point
\usepackage{bm}% bold math
\usepackage{hyperref}% add hypertext capabilities
\usepackage{upgreek}
\usepackage{color}
\usepackage{soul}
\usepackage{epstopdf}
\usepackage{units}
\usepackage{longtable}
\usepackage{floatrow}
\usepackage{latexsym}
\usepackage{gensymb}
\usepackage{floatrow}
\usepackage{mhchem}

\begin{document}

\title{Quantum anomalous Hall edge channels survive up to the Curie temperature}

\author{Kajetan M. Fijalkowski}
\email[email:]{kajetan.fijalkowski@physik.uni-wuerzburg.de}
\author{Nan Liu}
\author{Pankaj Mandal}
\author{Steffen Schreyeck}
\author{Karl Brunner}
\author{Charles Gould}
\email[email:]{charles.gould@physik.uni-wuerzburg.de}
\author{Laurens W. Molenkamp}
\affiliation{Faculty for Physics and Astronomy (EP3), Universit\"at W\"urzburg, Am Hubland, D-97074, W\"urzburg, Germany}
\affiliation{Institute for Topological Insulators, Am Hubland, D-97074, W\"urzburg, Germany}

\date{\today}

%\begin{abstract}
%\end{abstract}

\maketitle

\section[Abstract]{Abstract}

Achieving metrological precision of quantum anomalous Hall resistance quantization at zero magnetic field so far remains limited to temperatures of the order of 20 mK, while the Curie temperature in the involved material is as high as 20 K.
The reason for this discrepancy remains one of the biggest open questions surrounding the effect, and is the focus of this article.
Here we show, through a careful analysis of the non-local voltages on a multi-terminal Corbino geometry, that the chiral edge channels continue to exist without applied magnetic field up to the Curie temperature of bulk ferromagnetism of the magnetic topological insulator, and that thermally activated bulk conductance is responsible for this quantization breakdown. 
Our results offer important insights on the nature of the topological protection of these edge channels, provide an encouraging sign for potential applications, and establish the multi-terminal Corbino geometry as a powerful tool for the study of edge channel transport in topological materials.

\section[Introduction]{Introduction}

The quantum anomalous Hall effect (QAHE), first discovered in Cr/V-doped (Bi,Sb)$_{2}$Te${_3}$ \cite{Yu2010,Chang2013,Chang2015}, has opened new avenues for academic studies into solid state manifestations of axion electrodynamics \cite{Nomura2011,PhysRevB.92.081107,PhysRevB.92.085113,tokura1,Grauer2017,PhysRevLett.120.056801} and unconventional magnetism \cite{Bestwick2015,Lachman2015,Grauer2015,Yasuda2016,Fijalkowski2020}. However, metrological precision of Hall resistance quantization at zero magnetic field so far remains limited to temperature of the order of 20 mK \cite{Goetz2018,Fox2018,Okazaki2020}, while the Curie temperature ($T_{\mathrm{C}}$) in the involved materials is as high as 20 K \cite{Chang2013,Chang2015}.

Given this discrepancy in the temperature scales, a question which has been around since the discovery of the QAHE pertains to the relevant energy scale that stabilizes the protection of the topological state. As the Hall resistance takes on values clearly below $h/e^2$ (where \textit{h} is Planck's constant and \textit{e} the elementary charge) at temperatures above 1 K or so, well below $T_{\mathrm{C}}$, where the bulk remains robustly ferromagnetic, the non-quantized Hall resistance can in principle originate from the ordinary bulk states in the absence of any chiral edge channel. Moreover, the possible presence of two distinct ferromagnetic phases \cite{Fijalkowski2020} and the otherwise rich magnetic behaviour reported at low temperature \cite{Bestwick2015,Yasuda2016,Liu2017,He2018,Jiang2020,Wang2021} leave open the possibility of a second phase transition being involved, and thus possibly playing a role in the edge channel formation at some temperature below $T_{\mathrm{C}}$.

Recently, experimental progress has been made in significantly increasing the zero magnetic field anomalous Hall resistance at somewhat higher temperatures \cite{Mogi2015,Ou2017} by modulating or mixing the magnetic dopants. However, these results are still at temperatures well below $T_{\mathrm{C}}$, and it remains unclear what energy scale drives these observations, as they do not result from a change in the band gap of the material, nor any other clearly relevant sample characteristic. Other previous reports have suggested that the chiral edge channels coexist with other conducting channels at higher temperatures. This includes investigations of different non-local measurement configurations \cite{Chang2015b}, and non-reciprocal signals \cite{Yasuda2020}, both in a traditional Hall bar geometry. The main challenge for analyzing any results obtained from a traditional Hall bar, is the fact that the signals measured at the voltage leads can result from the current carried by both the edge channels and the bulk states in the material, making it impossible to definitively rule out contribution from the ordinary anomalous Hall effect.

Another established transport geometry is a two-terminal Corbino ring. The ring shape allows for circulating current components, and provides access to additional bulk transport properties \cite{Mumford2020,Kavokin2020}. In the context of topological materials with edge channels, the obvious limitation for such traditional Corbino setting is the absence of material edges extending directly between both contacts, making such measurement sensitive solely to the conductance in the bulk, and insensitive to the existence of edge channels.

In this work, we propose a solution that combines the two geometries, and opens up the possibility of directly probing the existence of chiral edge channels regardless of bulk conductance in the sample. A key strategy to distinguish between the quantum anomalous Hall effect, and the anomalous Hall effect of the bulk ferromagnet, is to spatially separate the current paths taken by bulk and edge channel transport contributions. Here, we achieve a clear separation of current paths using a non-local measurement scheme in a novel multi-terminal Corbino geometry. The measured signals directly reveal that the quantum anomalous Hall edge channels survive up to the $T_{\mathrm{C}}$ of bulk ferromagnetism in a magnetic topological insulator.

\section[Results]{Results and Discussion}

\subsection[The multi-terminal Corbino device]{The multi-terminal Corbino device}

The multi-terminal Corbino device is presented in Fig. 1a. The structure is patterned from a 8.2 nm thick V$_{0.1}$(Bi$_{0.2}$Sb$_{0.8}$)$_{1.9}$Te${_3}$ layer grown by molecular beam epitaxy (MBE) \cite{Winnerlein2017}, and optimized for exact anomalous Hall resistance quantization \cite{Goetz2018}. The outer diameter of the ring is 1 mm, the width between the edges is 100 $\mu$m, and the width of each contact is 15 $\mu$m. Four AuGe ohmic contacts along each the outer and the inner edge allow for four-terminal measurements along the same edge in the low temperature perfectly quantized state, and also for various measurement configurations involving the two edges. These allow a clear determination of the role of edge channels. The sample is also fitted with an AlOx/HfOx/Au top gate for tuning of the Fermi level. The width of 100 $\mu$m between the edges is large enough to clearly decouple the chiral edge channels residing on both edges, when the bulk is insulating. This is empirically verified by measurements demonstrating perfect quantization at low temperature (see Fig. 1).

\subsection[Basic characterization]{Basic characterization}

We first consider a two-terminal measurement applying a voltage between contact A and contact 1 (which is grounded), and measuring the flowing current with a series resistor. Such data is presented in Fig. 1b at a temperature of 25 mK as a function of gate voltage. For gate bias values in the QAHE plateau region (with the center of the plateau at about +5.5 V), the current, and thus the conductance vanishes. Indeed, below 55 mK the two-terminal resistance $V_{\mathrm{A-1}}$/$I_{\mathrm{A-1}}$ is noise limited (by the noise on the voltage drop over a series reference resistor) with a lower bound estimate of 11.2 G$\Omega$. Given the high conductance of edge channels running along each edge, in comparison to the bulk conductance at such low temperatures, equipotential rings should occur on the outer and the inner edges of the Corbino. This is confirmed experimentally by measuring the potential at the remaining contacts, which for each edge are all equal to each other to well within the 1 $\%$ experimental uncertainty on the lock-in amplifiers. For the purposes of determining the bulk resistivity, we effectively then have a Corbino with one inner and one outer contact, and the aspect ratio entering the calculation is given by the width of the ring divided by its circumference.  This ratio is approximately 1/28, yielding a  lower bound  of 310 G$\Omega$ per square for the bulk sheet resistivity ($\rho_{\mathrm{Bulk}}$). That such an extremely high resistivity can be measured in a cryostat is a first example of the usefulness of the multi-terminal Corbino geometry. The Fig. 1c shows a thermal activation plot of bulk resistivity with an Arrhenius activation energy corresponding to a temperature ($\Delta$ in the Fig. 1c) of 1.11 K. 

%----------------------------------------------------------------------------FIG1
\begin{figure}[t]
\includegraphics[width=\columnwidth]{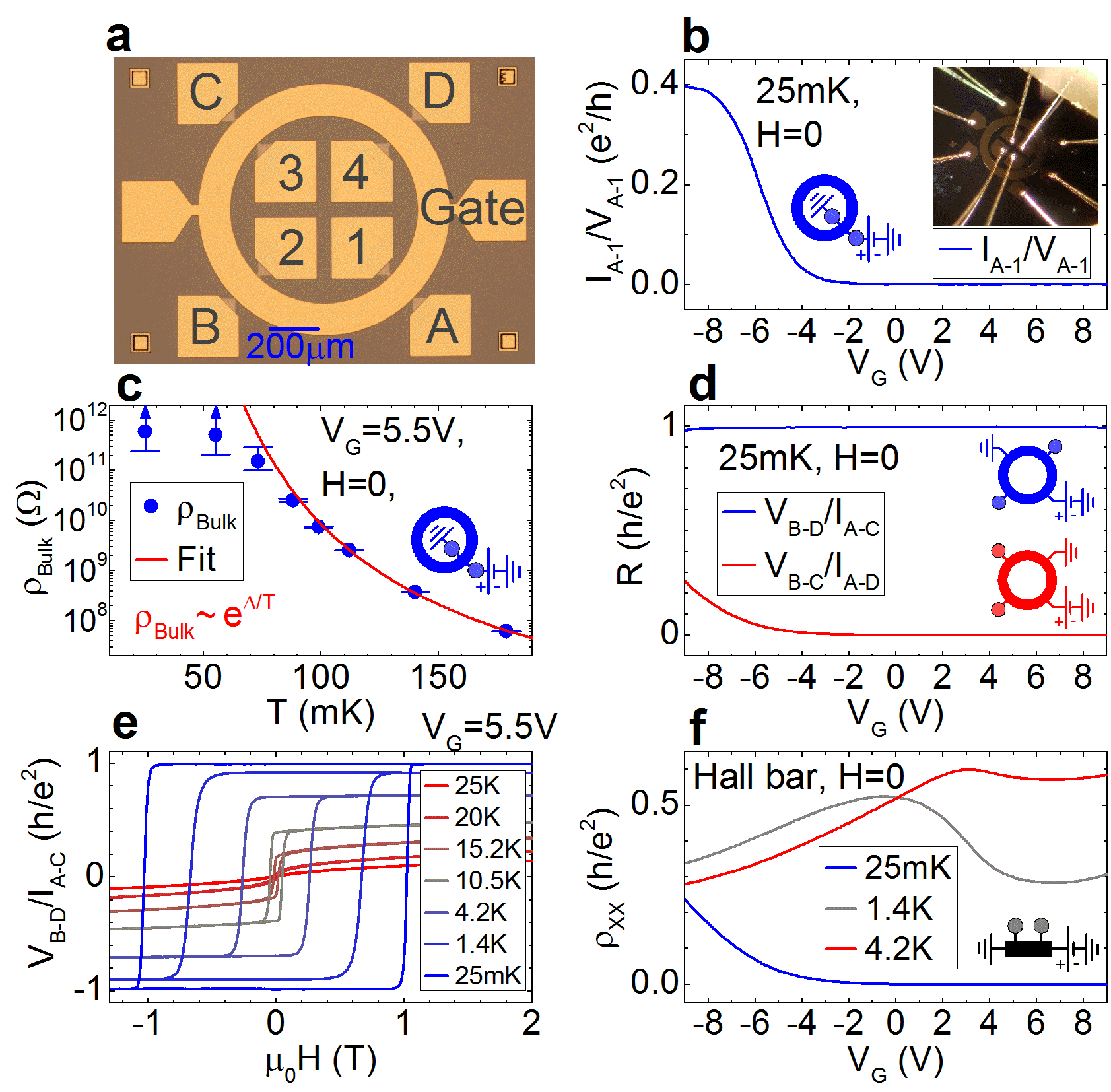}
\caption{a) Optical microscope image of the multi-terminal Corbino device with labeled ohmic contacts on the inner edge (1-4), on the outer edge (A-D), as well as a top gate. The smaller image in (b) shows a photograph of a mechanically bonded device. b) A base temperature two-terminal conductance gate sweep measurement between contacts A and 1 (between the two edges). c) A thermal activation of the bulk resistivity, with the red line representing an Arrhenius fit with an activation energy of 1.11 K. The upper and lower value for the error bars are determined by one standard deviation on the signal measured at the reference resistor. At 25 mK and 55 mK the upper value is not determined as the standard deviation exceeds the mean value of the signal (schematically indicated by the blue arrows). d) Four terminal measurements on the outer edge, at base temperature, showing perfect dissipationless chiral edge channel transport. e) External magnetic field hysteresis loop of $V_{\mathrm{B-D}}$/$I_{\mathrm{A-C}}$ for temperatures ranging from 25 mK to 25 K. f) Longitudinal resistivity measurements collected from an ordinary Hall bar device (with width of 200 $\mu$m and aspect ratio 3:1), patterned from the same MBE layer as the Corbino device.
}
\label{fig:Fig1}
\end{figure}
%---------------------------------------------------------------------------

%----------------------------------------------------------------------------FIG2

\begin{figure*}[]
\includegraphics[width=\columnwidth]{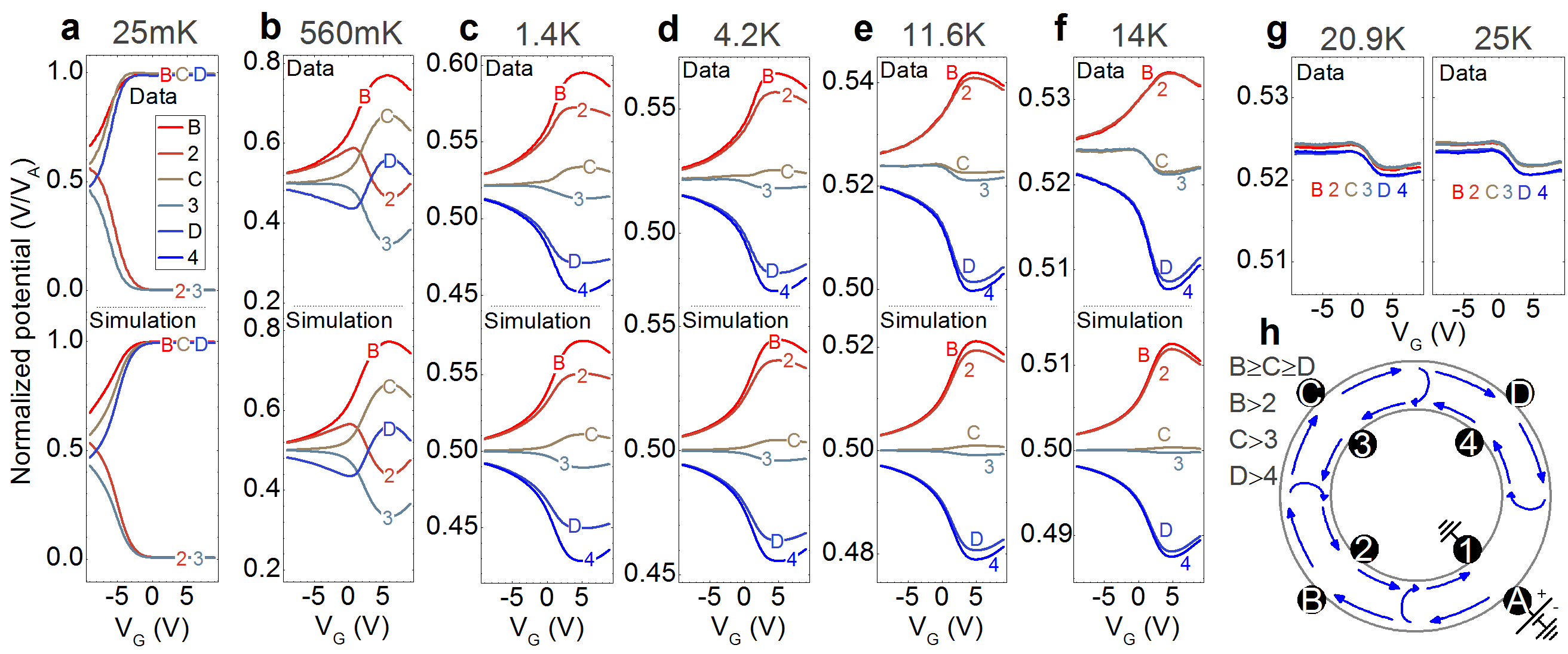}
\caption{Individual contact potentials in the magnetized state at zero external magnetic field, normalized to the bias voltage. The bias voltage is applied to contact A with contact 1 grounded. The temperature of the measurement is: 25 mK (a), 560 mK (b), 1.4 K (c), 4.2 K (d), 11.6 K (e), 14 K (f), and 20.9 K with 25 K (g). Below the data in a-f, the corresponding simulated curves are plotted. (Contact 4 was not accessible during the measurement run below 1.4 K). At 1.4 K and above, the measurements of contacts B and 2 were multiplied by a factor of 1.00357, and contacts C and 3 by a factor of 0.9943, to correct for the amplifier gain differences between the different instruments in the experiment. Color coding is the same in (a-g). h) A schematic of the system's chirality, with a description of the contact voltage order stemming from the chiral edge channel transport with bulk dissipation.
}%
\label{fig:Fig2}%
\end{figure*}

%---------------------------------------------------------------------------

When four-terminal measurements are performed at 25 mK along a single edge, perfectly dissipationless chiral edge channel transport is observed. In Fig. 1d we plot the $V_{\mathrm{B-D}}$/$I_{\mathrm{A-C}}$ (blue color) and the $V_{\mathrm{B-C}}$/$I_{\mathrm{A-D}}$ (red color) resistances, revealing similar characteristics to $R_{\mathrm{xy}}$ and $R_{\mathrm{xx}}$ of a perfectly quantized state in a traditional Hall bar, where $V_{\mathrm{B-D}}$/$I_{\mathrm{A-C}}$ reaches a quantized value of $h/e^{2}$ and $V_{\mathrm{B-C}}$/$I_{\mathrm{A-D}}$ vanishes. 

In Fig. 1e we show a magnetic field sweep of the $V_{\mathrm{B-D}}$/$I_{\mathrm{A-C}}$ signal for temperatures from 25 mK to 25 K. This configuration corresponds to an $R_{\mathrm{xy}}$ in a traditional Hall bar. At 25 mK (blue color) the curve has a rectangular-shaped hysteresis loop and shows perfect quantization. When the temperature is increased the signal decreases, and at temperatures around 18 K, the Curie temperature of the film, the zero magnetic field Hall signal vanishes. 

Fig. 1f shows longitudinal resistivity $\rho_{\mathrm{xx}}$ obtained from a Hall bar patterned from the same MBE layer. The $\rho_{\mathrm{xx}}$ value measured here results from the combined effect of the bulk resistivity and the edge channel contribution. This can lead to a non-monotonicity as seen most clearly at 1.4 K, which reflects the competing effects of bulk resistivity increasing as the Fermi level enters the bulk gap, and the dissipationless edge channel shorting the two voltage contacts when backscattering through the bulk is suppressed.   

\subsection[Non-local measurements]{Non-local measurements}

To gain insight on the role of edge channels as a function of temperature, we turn to non-local measurements involving both edges of the Corbino. Specifically, we apply a bias voltage to contact  A with contact 1 grounded, and record the individual potentials (with reference to the ground) at the remaining contacts, as a function of gate voltage and temperature. 

In the top half of each panel in Fig. 2, we plot zero-magnetic-field gate voltage sweeps for this configuration, at various temperatures. In all cases, the potentials are normalized to the voltage applied at contact A (typically 100 $\mu$V below 1 K, and 5 mV at higher temperatures). In the low temperature regime, when the system is tuned into the QAHE plateau (around +5 V), we observe that contacts along the outer edge are equipotential to the biased contact while those along the inner edge are equipotential to ground, consistent with pure chiral edge channel transport and a fully insulating bulk. 

Richer behavior is observed when bulk conductivity is turned on, either by tuning the Fermi level with gate voltage or by increasing the temperature. For all temperatures, including 25 mK where perfect quantization is observed, at a gate voltage of -9V, the Fermi level is far away from the charge neutrality point, and the bulk is thus conducting. As part of the current can now directly flow from contact A to 1, the equipotential situation along the outer and inner edges vanishes, and the potentials at the remaining contacts take on values closer to 0.5, the average potential between A and 1. Noteworthy however is that a clear chirality is observed in the data with the potential $V_{\mathrm{B}}$$>$$V_{\mathrm{C}}$$>$$V_{\mathrm{D}}$ and $V_2$$>$$V_3$$>$$V_4$ for all temperatures. This shows that even though the bulk is conducting, chiral edge channel transport remains readily accessible.

We now examine a gate voltage of +5 V, where the Fermi-level is in the bulk gap, and pure edge channel transport is observed at low temperatures. As the temperature is increased, the conductivity of the bulk is progressively turned on. This gradually changes the ratio of edge channel to bulk conductivity, and causes the potential at the detection contacts to move towards the average value of 0.5 (note that  for better visibility of the data, the y axis changes between the various temperatures in Fig. 2). Also in this case, the above described chirality is observed.   

The situation only changes in panel 2g, which shows the behavior of the potentials above the 18 K Curie temperature of the material. Here all contacts are, to within experimental error, equipotential to each other. The value of 0.52 instead of 0.5 results from a slight difference in contact resistance between contacts A and 1, and the change in this value around 0 V gate voltage comes from the effect of the gate on the resistance of these contacts.  

%----------------------------------------------------------------------------FIG3
\begin{figure}
\includegraphics[width=\columnwidth]{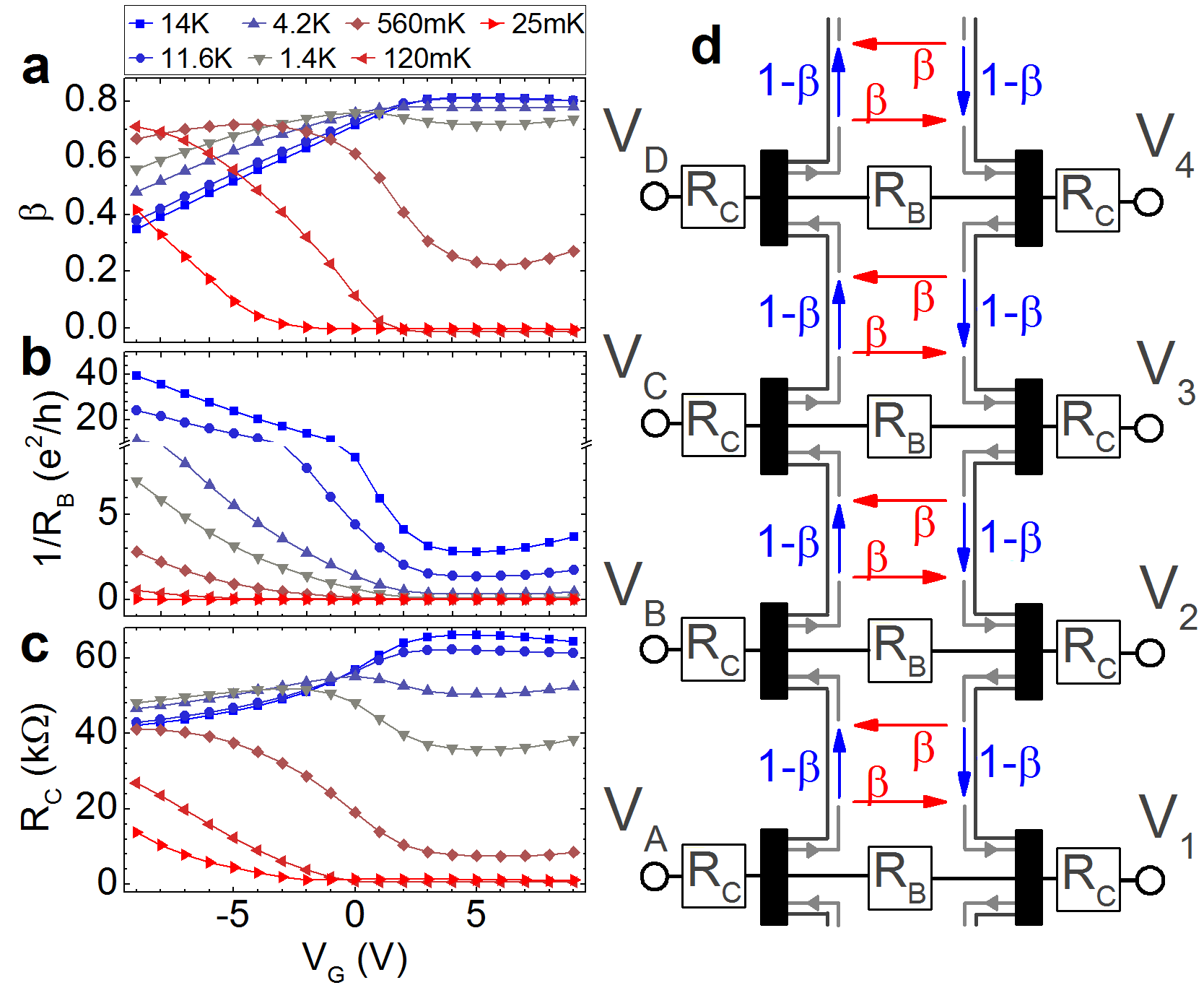}
\caption{Gate voltage and temperature evolution of the fitting parameters (see description in the text): a) $\beta$, b) 1/$R_{\mathrm{B}}$, and c) $R_{\mathrm{C}}$. d) Schematic of the device layout describing the model. The red and blue arrows represent the transmission coefficients dependent on $\beta$. Resistors $R_{\mathrm{C}}$ simulate the lead resistance in series with every contact, and resistance $R_{\mathrm{B}}$ the inverse bulk conductance. 
}%
\label{fig:Fig3}
\end{figure}
%---------------------------------------------------------------------------

%----------------------------------------------------------------------------FIG4
\begin{figure}
\includegraphics[width=\columnwidth]{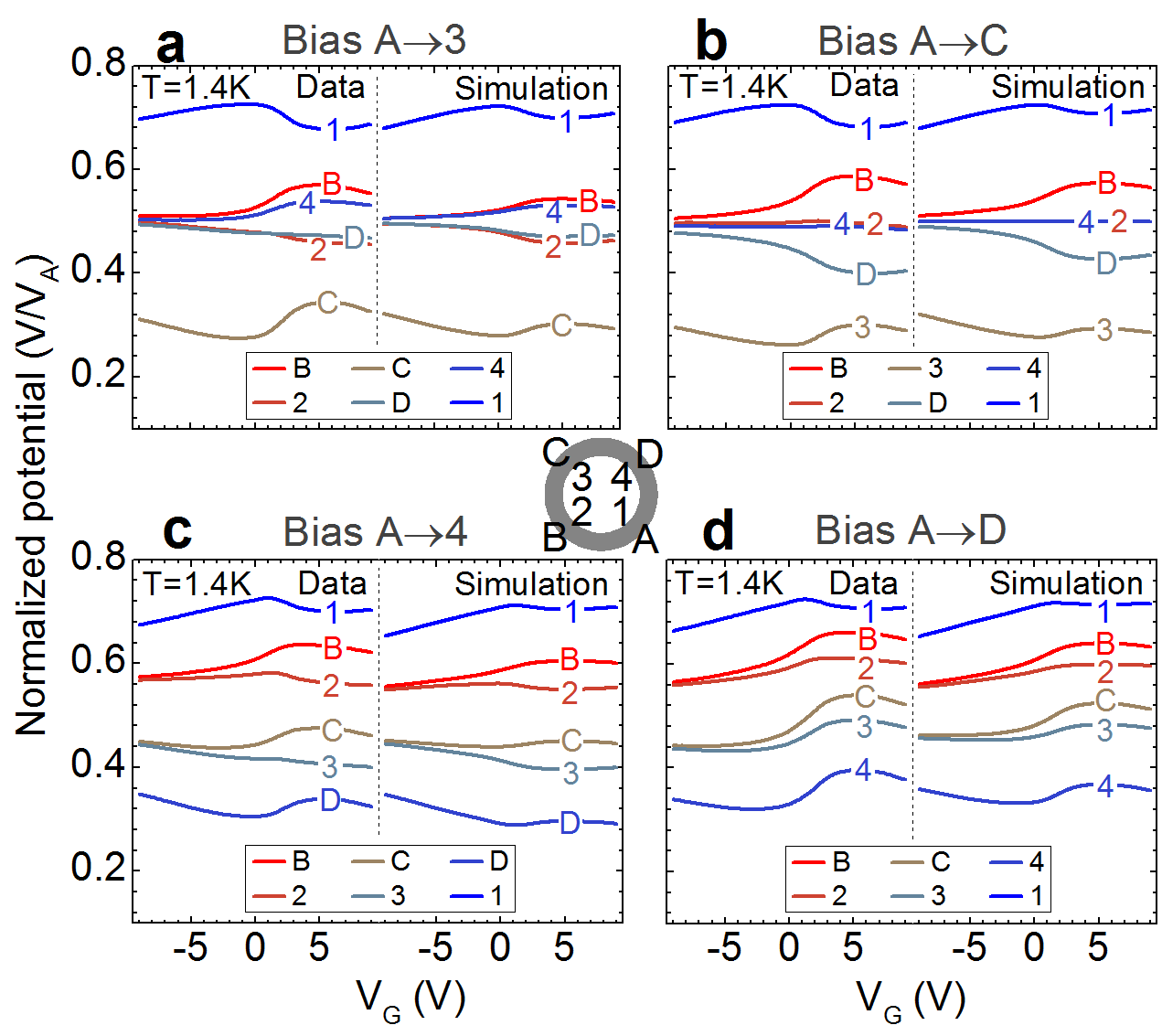}
\caption{Side by side comparison of the experimentally measured individual potentials in the magnetized state at zero external magnetic field (left in each figure), and simulated using the model (right in each figure), normalized to the bias voltage. All curves were collected at 1.4 K. Biased contact pairs (source-drain) in the figures are: A-3 (a), A-C (b), A-4 (c), and A-D (d).
}
\label{fig:Fig4}
\end{figure}
%---------------------------------------------------------------------------

We qualitatively interpret these data as follows: In the absence of any edge channels, one would expect a fairly direct potential gradient from contacts A to 1. This highlights the key difference from a typical Hall bar geometry, which allows us to rule out any bulk anomalous Hall contribution to the observed voltages. The origin of the anomalous Hall effect of a bulk ferromagnet is completely distinct from the quantum anomalous Hall effect. It stems from the intrinsic and extrinsic mechanisms resulting in asymmetric scattering of current carrying electrons, and thus only plays a role where current is flowing. Since, in contrast to the case of a Hall bar, no bulk current flows near the voltage leads in the Corbino geometry, the anomalous Hall effect has no effect on our measurements. The potentials resulting from bulk transport, measured at the positions of all voltage contacts, are therefore very close to 0.5, and in particular $V_{\mathrm{B}}$=$V_{\mathrm{D}}$ ($V_2$=$V_4$) due to symmetry.

The presence of chiral quantum anomalous Hall edge channels leads to an actual current flow as depicted in Fig. 2h. Current exiting contact A moves clockwise (for the given magnetization) around the outer edge, and is also able to partly backscatter through the bulk to the inner edge channel where it moves counter-clockwise. This produces both the observed chirality $V_{\mathrm{B}}$$>$$V_{\mathrm{C}}$$>$$V_{\mathrm{D}}$ as well as the finite voltage drops between outer and inner edge at all points around the ring. Of course, the opposite chirality is observed for reversed magnetization. Above 18 K, the QAHE edge channels disappear, and we recover the case corresponding to only a conducting bulk. 

\subsection[Landauer-B{\"u}ttiker modelling]{Landauer-B{\"u}ttiker modelling}

To further quantify our interpretation, we use a simple model based on the Landauer-B{\"u}ttiker formalism \cite{Buttiker1988}. The model is depicted in the schematic in Fig 3d and consists of an 8-contact Landauer-B{\"u}ttiker network to describe the chiral edge channels. The role of the conducting bulk is accounted for in two ways: by a set of resistors $R_{\mathrm{B}}$ linking each pair of inner and outer contacts to account for bulk conductivity (treated in the formalism as an additional transmission channel), and by a scattering parameter $\beta$ that allows for backscattering between the outer and the inner edge channels at ring positions between the contacts. Given the much larger distances involved, bulk conductivity between non-neighboring contacts can be safely neglected. This network leads to 8 linear equations with two free parameters ($\beta$ and $R_{\mathrm{B}}$) that can be solved analytically for any combination of current and voltage contacts. 

The final element needed to completely describe the data is $R_{\mathrm{C}}$, which is the contact/lead resistance at each of the 8 ohmic contacts. This $R_{\mathrm{C}}$ primarily stems from the sheet resistance of the mesa constriction between the metallic contact and the ring. This minimalistic model with only 3 adjustable fitting parameters is sufficient to describe all measurements. These three parameters can be jointly obtained by fitting the model to the following three measurements: The two-terminal resistance $V_{\mathrm{A-1}}$/$I_{\mathrm{A-1}}$, the two terminal resistance $V_{\mathrm{A-C}}$/$I_{\mathrm{A-C}}$, and the four terminal (effective $R_{\mathrm{xy}}$) resistance $V_{\mathrm{B-D}}$/$I_{\mathrm{A-C}}$ (see the Methods section for more details). Once the three parameters are obtained, they are used to generate the simulations shown in the bottom panels of Fig. 2, for our main measurement configuration. We emphasise that no fitting to the data in Fig. 2 was performed, as the model is already fully determined by the three above listed measurements for any given temperature and gate voltage.

The actual parameters used to generate the simulations in Fig. 2 are given in Fig. 3a-c as a function of gate voltage and temperature. The dependence of the contact/lead resistance $R_{\mathrm{C}}$ of Fig. 3c on both temperature and gate voltage matches that of the total mesa resistivity of the material as obtained from measurements on a Hall bar (shown in Fig. 1f). Note that this resistance results from the parallel conductance of the bulk and an edge channel, and is thus distinct from the pure bulk resistance $R_{\mathrm{B}}$ of Fig. 3b.  The inverse bulk resistance (1/$R_{\mathrm{B}}$) also follows the expected behavior, with conductivity turning on with either temperature or as gate voltage is tuned away from the plateau regime. The dependence of the parameter $\beta$ (Fig. 3a) roughly correlates with the bulk conductivity, consistent with the bulk mediating backscattering between the two edges. The somewhat non-monotonic behavior may result from the fact that when the bulk gets highly conducting, backscattered carriers begin to find pure bulk paths to the other contacts, which is not accounted for in our basic model. Note that a value of $\beta=0$ represents ideal edge channel conduction with a fully insulating bulk, whereas $\beta=1$  would be tantamount to a complete destruction of the edge channel. Values in between are a measure of the relative weight of bulk vs edge channel conduction.

Note that the existence of a contact resistance $R_{\mathrm{C}}$ also limits the degree of chiral splitting observable in the experimental data of Fig. 2. For example the spread from 0.53 to 0.51 between contacts B and D at 14 K. The asymmetry between B and D produced from the Corbino contribution itself, is thus larger than the 2 \% seen in the figure. This was confirmed with measurements on a second sample (Supplementary Note 1 and Supplementary Fig. 1) with wider leads (50 $\mu$m) to minimize the contact resistance. It has a splitting of approx. 8 \% under comparable conditions.  

The robustness of our model can be further tested by comparing it to additional measurement configurations, again without introducing any new fitting parameters. A sampling of other source-drain contact arrangements is presented in Fig. 4. In all cases, good agreement between the model and the data confirms the existence of chiral edge channel transport. 

Our observations thus clearly show that the quantum anomalous Hall chiral edge channels survive over a temperature range vastly exceeding the regime where quantization is directly observable in experiment. This implies that the breakdown mechanism for this quantization is not a feature of the topological edge channels, but rather results from parasitic conductance through a non-fully-insulating bulk. This in turn suggests that the topological protection in these systems is robust against temperature, and that efforts to increase operation temperatures for potential future applications should focus on optimizing the insulating properties of the bulk. In addition, these results establish the multi-terminal Corbino geometry as a powerful experimental tool for studies of edge channel transport in topological materials.

\bigskip

\section[Methods]{Methods}

\subsection[Device preparation]{Device preparation}
Our magnetic topological insulator V$_{0.1}$(Bi$_{0.2}$Sb$_{0.8}$)$_{1.9}$Te${_3}$ layer is grown using molecular beam epitaxy (MBE) on a Si(111) substrate \cite{Winnerlein2017}. The layer is capped in-situ with an 8 nm thick insulating Te layer as a protection from ambient conditions as well as from chemicals in the lithographic process. The magnetic TI layer thickness is determined using X-ray reflectivity (XRR) measurements to be approximately 8.2 nm. A multi-terminal Corbino geometry is patterned using standard optical lithography methods, with AuGe ohmic contacts and an AlOx/HfOx/Au top gating layer stack allowing for tuning of the Fermi level.

\subsection[Transport measurements]{Transport measurements}
Transport measurements are performed in a \ce{^{3}He}-\ce{^{4}He} dilution refrigerator system (for measurements below 1 K) and in a \ce{^{4}He} cryostat (for measurements above 1 K), with an out of plane external magnetic field. Measurements are done in the linear response regime using low frequency AC voltage excitation (below 20 Hz), with the exception of the bulk resistivity plotted in Fig. 1c where a quasi-DC square wave (approx. 4 mHz) excitation is used instead.

\subsection[Method of obtaining the model parameters]{Method of obtaining the model parameters}
The three parameters ($\beta$, $R_{\mathrm{B}}$, $R_{\mathrm{C}}$) can be obtained by matching the three experimentally measured resistances ($R_{\mathrm{A1,A1}}$=$V_{\mathrm{A-1}}$/$I_{\mathrm{A-1}}$, $R_{\mathrm{AC,AC}}$=$V_{\mathrm{A-C}}$/$I_{\mathrm{A-C}}$, $R_{\mathrm{BD,AC}}$=$V_{\mathrm{B-D}}$/$I_{\mathrm{A-C}}$) to the same resistance configurations calculated from the Landauer-B{\"u}ttiker network. By analytically solving the Landauer-B{\"u}ttiker circuit for these configurations of current and voltage leads, one obtains the following equations:

\begin{gather}
 R_{\mathrm{A1,A1}}
 =
 \frac{h}{e^2}
 \frac{3\beta+1+4/R_{\mathrm{B}}}{4(\beta+1/R_{\mathrm{B}})(1+1/R_{\mathrm{B}})}
 +2R_{\mathrm{C}}
\end{gather}
\begin{gather}
 R_{\mathrm{AC,AC}}
 =
 \frac{h}{e^2}
 \frac{1}{1-\beta}
 +2R_{\mathrm{C}}
\end{gather}
\begin{gather}
 R_{\mathrm{BD,AC}}
 =
 \frac{h}{e^2}
 \frac{1}{1+1/R_{\mathrm{B}}}
\end{gather}
which form a system of independent equations relating the three model parameters ($\beta$, $R_{\mathrm{B}}$, $R_{\mathrm{C}}$) to the three measured resistances ($R_{\mathrm{A1,A1}}$, $R_{\mathrm{AC,AC}}$, $R_{\mathrm{BD,AC}}$).

\bigskip

\section[Acknowledgents]{Acknowledgents}
We gratefully acknowledge the financial support of the Free State of Bavaria (the Institute for Topological Insulators), Deutsche Forschungsgemeinschaft (SFB 1170, 258499086), W\"urzburg-Dresden Cluster of Excellence on Complexity and Topology in Quantum Matter (EXC 2147, 39085490), and the European Commission under the H2020 FETPROACT Grant TOCHA (824140).
\bigskip

\section[Author contributions]{Author contributions}
K. M. F. designed and patterned the devices, performed the transport experiments, and the Landauer-B{\"u}ttiker modeling. N. L. grew the sample. S. S. contributed to material growth optimization, and P. M. to lithography optimization and material characterization. K. B., C. G., and L. W. M. supervised the work. All authors contributed to the analysis and interpretation of the results, and the writing of the manuscript.
\bigskip

\setcounter{figure}{0} 
\renewcommand{\figurename}{Supplementary Fig. }

\bigskip

\huge
\centerline{\textbf{Supplementary material}}
\normalsize

\section{Supplementary Note 1 - Non-local measurements on a multi-terminal Corbino device with wider contacts.}
Supplementary Fig. 1 shows non-local measurements collected from a second device, where the contact width is increased from 15 $\mu$m (for the sample from the main text) to 50 $\mu$m. A bias voltage of 5 mV is applied to contact A, while contact 1 is grounded, and voltages on all of the remaining contacts are measured with reference to the ground. All measured potentials are normalized to the voltage measured at contact A. The potentials are measured as a function of the applied gate voltage, at zero external magnetic field in a magnetized state, at temperatures: 1.4 K (a), 4.2 K (b), 11.6 K (c), 15.2 K (d), and above the Curie temperature, at 20 K (e).

The model parameters for this sample are obtained in the identical way as described in the main text. The corresponding simulated curves are presented in the bottom panels in (a)-(e), and good agreement is found between the data and the simulation. The evolution of the extracted model parameters with gate voltage and temperature is presented in (f). The behavior of the inter-edge scattering $\beta$ and the bulk conductance 1/$R_{\mathrm{B}}$ is consistent between both samples, and the contact resistance $R_{\mathrm{C}}$ is clearly reduced for the sample with wider contacts. This is consistent with all of the analysis, as the only difference in the device geometry between the two samples is the contact width [see an optical microscope image in (g) for the sample with wider contacts]. The smaller contact resistance leads to a significant increase in the magnitude of the non-local signals, with voltage splitting between contacts B and D of about 8 $\%$ at 15.2 K, right below the Curie temperature (which is approx. 18 K).

\onecolumngrid
\bigskip

%----------------------------------------------------------------------------FIGS1
\begin{figure}[h]
\includegraphics[width=15cm]{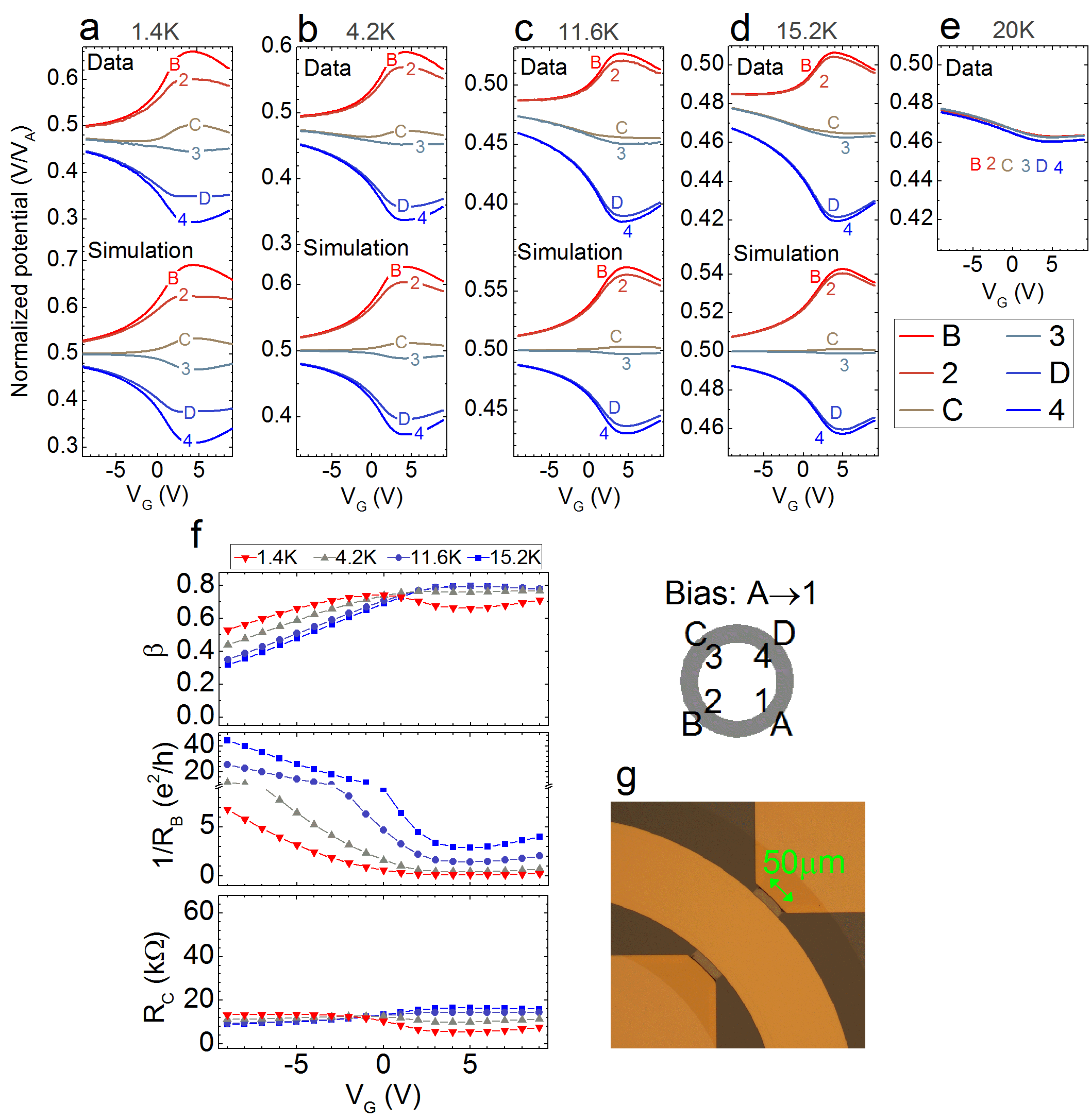}
\caption{
Individual contact potentials collected in the magnetized state at zero magnetic field, from a second sample with wider contacts (50 $\mu$m) at a variety of temperatures. Contact A is biased, and contact 1 is grounded. All potentials are normalized to the bias voltage. The measurements of contacts B and 2 were multiplied by a factor of 1.00357, and contacts C and 3 by a factor of 0.9943, to correct for the amplifier gain differences between the different instruments used in the measurements. Color coding is the same in (a-e). f) Corresponding modeling parameters obtained for this sample, using the same methodology as in the paper, as a function of gate voltage. g) A microscope image showing the widened contacts.
}
\label{fig:FigS1}%
\end{figure}
%---------------------------------------------------------------------------

\end{document}